\documentclass[aps,ams,amsmath,prx,superscriptaddress,longbibliography,twocolumn]{revtex4-1}
\usepackage{graphics}
\usepackage{graphicx}
\usepackage{epstopdf,epsfig}
\usepackage{amsmath}
\usepackage{amssymb}
\usepackage{amsfonts}
\usepackage{mathrsfs}
\usepackage{bm}
\usepackage{xcolor}
\usepackage{bbm}
\usepackage{hyperref}
\usepackage[normalem]{ulem}
\usepackage{comment}
\usepackage{soul}
\usepackage[varg]{txfonts}

\begin{document}

\title{Fluctuation-induced magnetoresistance in graphene Hall bars at charge neutrality}

\author{A. Grigoreva}
\affiliation{Department of Physics, University of Washington, Seattle, Washington 98195, USA}

\author{A. Levchenko}
\affiliation{Department of Physics, University of Wisconsin-Madison, Madison, Wisconsin 53706, USA}

\author{E. Kirkinis}
\affiliation{Center for Computation and Theory of Soft Materials, Robert R. McCormick School of Engineering and Applied Science, Northwestern University, Evanston IL 60208 USA}

\author{A. V. Andreev}
\affiliation{Department of Physics, University of Washington, Seattle, Washington 98195, USA}

\date{\today}

\begin{abstract} 
In the temperature interval  $T\sim 50 - 100 K$, the rate of electron-electron collisions in graphene devices may exceed the momentum relaxation rate due to disorder and electron-phonon scattering. In this regime, the motion of the electron liquid may be described by the hydrodynamic equations. In the hydrodynamic approximation, at the charge neutrality point the electric current is decoupled from the hydrodynamic flow, and the conductivity of the system is equal to the intrinsic conductivity of the electron liquid, $\sigma_0$. It was recently shown that the coupling of charge transport and hydrodynamic flow induced by the thermal fluctuations enhances the macroscopic conductivity of the system and results in very strong positive magnetoresistance (MR) at relatively weak fields. Here we develop a quantitative theory of the fluctuation mechanism of MR at charge neutrality and obtain analytic expressions for MR in graphene Hall bars.  The functional dependence of MR on the magnetic field $H$ is sensitive to the Hall bar geometry and  orientation relative to the current.  We also account for the effect of momentum relaxation on MR. 
\end{abstract}

\maketitle

\section{Introduction}\label{sec:intro}

The conduction electrons in graphene form a Dirac liquid \cite{Narozhny2017,Lucas2018,Narozhny2022,Fritz2024} characterized by a vanishing Fermi surface and linear band dispersion with velocity $v\approx 10^6\,\mathrm{m/s}$. 
Among the interesting properties of the Dirac liquid is its intrinsic conductivity $\sigma_0$. In typical experimental devices, the dimensionless coupling constant for the electron-electron interaction is not small, $\alpha = e^2/\varepsilon \hbar v \sim 1$ (we use Gaussian units). Here $e$ is the electron charge, and $\varepsilon$ is the dielectric constant of the substrate (for a typically used hexagonal boron nitride $\varepsilon \approx 3.4$, and $\alpha \approx 0.64$). This
makes a microscopic theory of the intrinsic conductivity a challenging problem~\cite{Mishchenko2007,SheehySchmalian2007,Kashuba2008,Fritz2008}.

The large value of the coupling constant $\alpha$ also complicates the microscopic analysis of the kinetics of the Dirac liquid at nonzero temperature. At $\alpha\sim 1$ the rate of inelastic collisions in a Dirac liquid reaches the so-called Planckian value $\hbar \tau_{\mathrm{ee}}^{-1} \sim T$, and the mean free path $l_{\mathrm{ee}}$ of the electrons is of the same order as the thermal de Broglie wavelength $\lambda_T =\hbar v/ T$ (the Ioffe-Regel limit). Thus, the Dirac liquid transitions from the collisionless quantum regime at high frequencies, $\hbar \omega > T $, directly to the collision-dominated hydrodynamic regime, $\omega \ll \tau_{\mathrm{ee}}^{-1}$, bypassing the regime of applicability of the Boltzmann kinetic equation.

An important feature of the electron hydrodynamics in Dirac liquids is the decoupling of the hydrodynamic flow from the electric current; the macroscopic motion of the liquid corresponds to the flow of heat, whereas the electric current is passed via the intrinsic conductivity of the electron fluid in the absence of the hydrodynamic flow \footnote{This aspect of electron hydrodynamics is shared by all compensated conductors and neutral plasmas.}. Because of this, 
the macroscopic conductivity of pristine graphene devices at the charge neutrality point was identified with the intrinsic conductivity of the electron liquid. 

It was recently shown~\cite{fluct_graphene} that this picture requires a revision. The reason is that the decoupling of the hydrodynamic flow from the electric current at charge neutrality holds only on average. Thermal fluctuations create deviations from local charge neutrality, causing coupling of the hydrodynamic flow and charge transport. 
This leads to the enhancement of the macroscopic conductivity of the system over the intrinsic conductivity $\sigma_0$. 

The mechanism of the fluctuation enhancement of the macroscopic conductivity may be qualitatively understood as follows. The Johnson-Nyquist current noise associated with the intrinsic conductivity $\sigma_0$ creates deviations of the local charge density from charge neutrality. The force exerted by the external electric field on the charge density fluctuations drives a fluctuating hydrodynamic flow, whose velocity field is correlated with the fluctuating charge density. In second order in the fluctuations, the advection of the charge density by this flow produces a nonvanishing contribution to the transport current, which increases the macroscopic conductivity of the system. 

The magnitude of the fluctuation enhancement of the macroscopic conductivity depends on the relaxation rate of charge fluctuations. In a typical experimental setup, where the long-range part of the Coulomb interaction is screened by the gate, the relaxation of charge density fluctuations proceeds by diffusion. In this case, for square samples of size $L$, the fluctuation enhancement of the macroscopic conductivity was shown to diverge logarithmically with $L$.
It was also shown in Ref.~\cite{fluct_graphene} that the fluctuation enhancement of the conductivity produces very strong, $\gtrsim 100\,\%$, magnetoresistance at weak fields, at which the magnetic length $l_H = \sqrt{\hbar c/|e|H}$ becomes of the order of $\sqrt{l_{\mathrm{ee}} L}$. In this regime, the intrinsic conductivity $\sigma_0$ is practically independent of $H$, and the MR is caused by the effect of the magnetic field on the spreading and relaxation of the hydrodynamic fluctuations.

\begin{figure}[t!]
    \centering
    \includegraphics[width=\columnwidth]{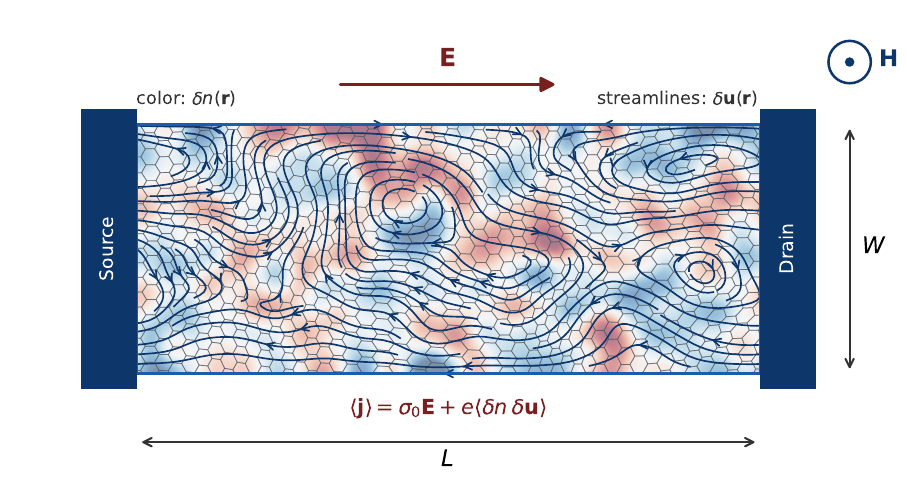}
    \caption{Illustration of the fluctuation enhancement of the macroscopic conductivity of a graphene Hall bar of length $L$ and width $W$ in a perpendicular magnetic field $H$. The color shows a snapshot of the thermal charge-density fluctuation $\delta n(\bm{r})$  generated by the Johnson–Nyquist current noise. The streamlines show the hydrodynamic flow $\delta\bm{u}(\bm{r})$ induced by the charge density fluctuations in the presence of a bias field $\bm{E}$. The correlated advection $e\langle\delta n\delta\bm{u}\rangle$ increases the transport current through the device.}
    \label{fig:Hall-bar}
\end{figure}

In the present work, we extend the theory of fluctuation-induced MR in graphene devices in the following directions: (1) we include the effect of fluctuations of the stress tensor associated with the viscosity of the electron liquid \footnote{In the treatment of the fluctuation conductivity in Ref.~\cite{fluct_graphene} the random stresses associated with the viscosity of the electron liquid were ignored, because at $H=0$ the charge fluctuations produced by the longitudinal component of the viscous force are subleading in the powers of the gradient expansion to the contribution of the Johnson-Nyquist noise. Here we show that the mixing of the longitudinal and transverse fluctuations that arises at $H\neq 0$ makes these stresses contribute to the MR at the same order as the current noise.}; (2) we extend the treatment of Ref.~\cite{fluct_graphene}, which focused on square samples, to Hall bars of width $W$ and length $L$, see Fig. \ref{fig:Hall-bar}, and show that the functional form of MR depends substantially on the aspect ratio $W/L$. This may help identify the fluctuation mechanism of MR in experiments; (3) we account for the effect of weak momentum relaxation caused by, e.g., disorder or phonon scattering. We also account for the fluctuating electrochemical potential in the EMF (second term in Eq.~\eqref{eq:EMF_def}), which was neglected in  Ref.~\cite{fluct_graphene}.

The presentation below is organized as follows. In Sec. \ref{sec:hydro} we introduce the Langevin description of thermal fluctuations of the electron liquid in the hydrodynamic regime at the charge neutrality point. In Sec. \ref{sec:conductivity} we evaluate 
the fluctuation magnetoconductivity of Hall bars with an arbitrary aspect ratio $W/L$. We discuss the results in Sec. \ref{sec:summary}. The details of the calculations are presented in the Appendix.

\section{Hydrodynamic fluctuations 
at charge neutrality}\label{sec:hydro}

We consider a two-dimensional electron liquid at charge neutrality in a perpendicular magnetic field $\bm H = H\hat z$, in the regime where electron-electron collisions equilibrate the system locally. The hydrodynamic description applies in the regime where the rate of 
momentum-conserving electron-electron collisions, $\tau_{\mathrm{ee}}^{-1}$, exceeds the momentum relaxation rate, $\gamma_p$, caused by scattering of electrons by disorder and phonons.  Furthermore, the inelastic mean free path $l_{\rm ee}\sim v \tau_{\mathrm{ee}} $  must be shorter than the system size and other length scales in the problem. 

The magnetic field is assumed classically weak, so that the magnetic length $l_H$ exceeds the microscopic scale corresponding to the onset of the hydrodynamic regime.  This assumption allows us to neglect the dependence of the thermodynamic quantities and kinetic coefficients of the liquid on the magnetic field~\cite{fluct_graphene,MullerSachdev2008}. For the Dirac electron liquid in graphene, this assumption is valid provided  $l_H \gg \lambda_T$.
Below, we set $\hbar = k_B = 1$ unless indicated otherwise.

The Langevin description of the fluctuations in the electron liquid in the hydrodynamic regime at the charge neutrality point is constructed by adding random Langevin sources to the dissipative currents and a random Langevin force to the momentum balance equation~\cite{LandauLifshitz1957,LandauV,LL9}. Because the liquid is in local thermal equilibrium, the variances of the Langevin sources are determined by the fluctuation-dissipation theorem (FDT).

The electric current density in the presence of fluctuations is given by
\begin{align}
    \label{eq:current}
    \bm{j} =  \sigma_0\, \bm{\mathcal{E}}  + e n \bm{u} +  \bm{\mathcal{J}}.
\end{align}
The first term on the right-hand side (RHS) is linear in the electromotive force (EMF) $\bm{\mathcal{E}}$ and describes the contribution of the intrinsic conductivity. The second term represents the advective part of the current caused by the fluctuations; in it, $n$ is the electron density, $\bm u$ is the hydrodynamic velocity, and $e$ is the electron charge. The third term is the random Langevin current whose variance is proportional to the intrinsic conductivity in accordance with the FDT,
\begin{align}
    \label{eq:jnoise_FDT}
    \langle\mathcal{J}_\alpha (\bm{r}_1, t_1)\, \mathcal{J}_\beta (\bm{r}_2, t_2) \rangle   =  2 T \sigma_0\, \delta_{\alpha \beta}\,
    \delta  (\bm{r}_1 -\bm{r}_2)\, \delta( t_1- t_2).
\end{align}
Here $\alpha$ and $\beta$ are Cartesian indices. In the presence of a static external electric field $\bm E$ and a magnetic field $\bm H$, the local fluctuating EMF is given by
\begin{align}
       \label{eq:EMF_def}
    \bm{\mathcal{E}} =  \bm{E} - \frac{1}{e} \bm{\nabla} \left(\mu + e \phi\right) + \frac{1}{c}\, [\bm{u}\times \bm{H}].
\end{align}
Here $\mu + e\phi$ is the fluctuating electrochemical potential of the electron liquid. The electric potential $\phi$ is related to the fluctuating electron density $n$ by the Poisson equation.

The macroscopic current density is obtained by averaging Eq.~\eqref{eq:current} over the fluctuations,
\begin{align}
    \label{eq:current_average}
    \langle \bm{j} \rangle =  \sigma_0\, \bm{E} + e \langle n \bm{u} \rangle .
\end{align}
The second term on the RHS represents the enhancement of the macroscopic current, which is caused by the advection of the electron density fluctuations by the fluctuating hydrodynamic flow. 

We show in Appendix \ref{sec:longitudinal} that the longitudinal fluctuations of the velocity produce a contribution to the macroscopic conductivity, which is smaller than the contribution of transverse velocity fluctuations in the parameter $\left(\tau_{\mathrm{ee}}\gamma_p\right)^2 \ll 1$. Therefore, in the main text, we focus our consideration on the transverse fluctuations of the hydrodynamic velocity. 

The transverse component of the linearized Navier-Stokes-Langevin equation has the form,
\begin{align}
    \label{eq:u_evolution}
    \rho\,\partial_t \bm{u}^{(t)} = & \,  e n \bm{E}^{(t)} + \frac{1}{c}\,[\bm{j}^{(l)}\times \bm{H}] - k\,\bm u^{(t)} \nonumber\\
    & \, + \eta \nabla^2 \bm{u}^{(t)}  + \bm f^{(t)} .
\end{align}
Here the superscripts $(t)$ and $(l)$ denote, respectively, the transverse and longitudinal components of the various vectors, $\rho$ is the mass density (defined as the coefficient relating the momentum density to the velocity $\bm{u}$), $\eta$ is the shear viscosity, and $k= \rho \gamma_p$ is the friction coefficient. Note that in two dimensions the cross product with $\bm H = H\hat z$ rotates in-plane vectors by $90^\circ$; hence only the longitudinal component $\bm j^{(l)}$ of the current produces a transverse Lorentz force, and, conversely, the transverse velocity generates a longitudinal contribution to the EMF.  
The fluctuating force density $\bm f = \bm f_k + \bm f_\eta$ arises from the random viscous stress tensor $\hat s$, $\bm f_\eta = \mathrm{div}\,\hat s$,  and the Langevin force associated with friction. Since the two contributions are statistically independent, using the Landau-Lifshitz result~\cite{LandauLifshitz1957,LL9} for the variance of $\hat s$,
we get 
the variance of the combined force:
\begin{equation}
\label{eq:fnoise_full}
\langle f_{\alpha}(\bm r, t) f_{\beta} (0, 0)\rangle =2 T \left(\delta_{\alpha\beta} \big[ k - \eta \nabla^2\big] - \zeta \nabla_{\alpha} \nabla_{\beta}\right) \delta(\bm r) \delta(t),
\end{equation}
where $\zeta$ is the bulk viscosity. Equation \eqref{eq:u_evolution} involves only the transverse component of this force. 

The density fluctuation in the first term on the RHS of Eq.~\eqref{eq:u_evolution} is related to the longitudinal component of the current by the continuity equation,
\begin{align}
    \label{eq:continuity_charge}
    e\, \partial_t n + \bm{\nabla} \cdot \bm{j}=0 ,
\end{align}
in which only $\bm j^{(l)}$ contributes to the divergence.

Equations \eqref{eq:current}-\eqref{eq:continuity_charge} constitute the full system of equations needed to describe the contribution of hydrodynamic fluctuations to the transport current through the system. Below, we solve these equations to determine the fluctuation contribution to the macroscopic conductivity of infinite systems and  Hall bar samples of length $L$ and width $W$. We also determine the magnetoresistance caused by the fluctuations.

\section{Fluctuation magnetoconductivity for Hall bar geometry}\label{sec:conductivity}

The fluctuation contribution to the current density is described by the second term on the RHS of Eq.~\eqref{eq:current_average} and arises in second order in the fluctuations. To evaluate it we solve the evolution equations for the nonuniform fluctuations of the electron density and the hydrodynamic velocity, Eqs.~\eqref{eq:current}, \eqref{eq:EMF_def}, \eqref{eq:u_evolution}, and \eqref{eq:continuity_charge}, to linear order in the Langevin sources and substitute the result into Eq.~\eqref{eq:current_average}.

We work in Fourier components and express the electrochemical potential fluctuations in terms of the density, $\mu_{\bm q,\omega} + e\phi_{\bm q,\omega} = [U(q) + \partial_n\mu]\, n_{\bm q,\omega}$. To linear order in the fluctuations, the longitudinal current at $\bm{q} \neq 0$ is
\begin{align}
\label{eq:jl}
\bm j^{(l)}_{\bm q,\omega} = & \, \sigma_0\left( -\frac{i\bm q}{e}\left[U(q) + \partial_n\mu\right]n_{\bm q,\omega} + \frac{1}{c}\,[\bm u_{\bm q,\omega}^{(t)} \times\bm H]^{(l)}\right) \nonumber\\
& \, + \bm{\mathcal{J}}^{(l)}_{\bm q,\omega},
\end{align}
Here $U(q)$ is the Fourier transform of the Coulomb interaction.

Substituting Eq.~\eqref{eq:jl} into Eq.~\eqref{eq:u_evolution}, multiplying the result by $\bm H\cdot[\bm q\times\ldots]$, and using the continuity equation \eqref{eq:continuity_charge}, we obtain a closed system of equations for the pair of variables $en_{\bm q,\omega}$ and $\bm H\cdot[\bm q\times\bm u^{(t)}_{\bm q,\omega}]$, 
\begin{subequations}
    \label{eq:sys}
\begin{align}
    & \left( - i \omega + \gamma_{\sigma} \right)e n_{\bm{q},\omega} + i\,\frac{\sigma_0}{c}\, \bm{H} \cdot [\bm{q} \times \bm{u}^{(t)}_{\bm{q},\omega}]   =   -  i\, \bm{q}\cdot  \bm{\mathcal{J}}^{(l)}_{\bm{q},\omega} , \label{eq:sysA}\\
    & \left(\bm{q} \cdot [\bm{E}^{(t)} \times \bm{H}] + i \gamma_{\sigma}\, \frac{H^2}{c}\right) e n_{\bm{q},\omega} \nonumber\\
    &\quad - \rho\left(- i \omega  +\gamma_\nu + \gamma_{p} + \gamma_H\right) \bm{H}\cdot [ \bm{q} \times  \bm{u}^{(t)}_{\bm{q},\omega}] \nonumber\\
    &\qquad =  \frac{H^2}{c}\, \bm{q}\cdot \bm{\mathcal{J}}^{(l)}_{\bm{q},\omega} + i \mathcal{F}_{\bm q,\omega}. \label{eq:sysB}
\end{align}
\end{subequations}
Here the various relaxation rates for fluctuations with a wavevector $q$ are denoted by $\gamma$ with a subscript; $\gamma_\sigma$ is the Maxwell relaxation rate proportional to the intrinsic conductivity $\sigma_0$, $\gamma_\nu$ is the viscous relaxation rate proportional to the kinematic viscosity $\nu =\eta/\rho$,  $\gamma_H$ is the relaxation rate associated with magnetic friction, and $\gamma_p$ is the momentum relaxation rate associated with the friction coefficient $k$. These rates are given by
\begin{align}
    \label{eq:rates}
   &\gamma_\sigma = \frac{\sigma_0}{e^2 } \left[U(q) + \partial_n \mu\right] q^2 , \qquad  \gamma_\nu = \nu q^2, \nonumber\\
  & \gamma_H = \frac{\sigma_0 H^2}{c^2 \rho}, \qquad  \gamma_p = \frac{k}{\rho}.
\end{align}

In Eq.~\eqref{eq:sysB} we introduced the notation 
\begin{align}
    \label{eq:F_def}
    \mathcal{F}_{\bm{q},\omega} = i \bm {H} \cdot \left[\bm{q} \times \bm{f}^{(t)}_{\bm{q},\omega} \right],
\end{align}
where $\bm{f}^{(t)}_{\bm q,\omega}$ is the Fourier amplitude of  the transverse Langevin force in Eq.~\eqref{eq:u_evolution}. The variance of the latter is obtained using Eq.~\eqref{eq:fnoise_full}  and is independent of  the bulk viscosity,
\begin{align}
\label{eq:fnoise_transverse}
\langle f^{(t)}_{\alpha}\, f^{(t)}_{\beta} \rangle_{\bm q,\omega} = 2 T \left( k + \eta q^2\right) \left(\delta_{\alpha\beta} - \hat q_\alpha \hat q_\beta\right) .
\end{align}
The Langevin noise correlators in the Fourier representation follow from Eqs.~\eqref{eq:jnoise_FDT} and \eqref{eq:fnoise_transverse},
\begin{subequations}
\label{eq:noise_q}
\begin{align}
\label{eq:noise_q_jj}
\langle \bm q\cdot\bm{\mathcal{J}}_{-\bm q,-\omega}\;\bm q\cdot \bm{\mathcal{J}}_{\bm q,\omega}\rangle &= 2T\sigma_0 q^2, \\
\label{eq:noise_q_ff}
\langle \mathcal{F}_{-\bm q,-\omega}\,\mathcal{F}_{\bm q,\omega}\rangle & = 2T\rho\,  (\gamma_\nu + \gamma_p) \, H^2 q^2, \\
\label{eq:noise_q_jf}
\langle \bm q\cdot\bm{\mathcal{J}}\;\mathcal{F}\rangle &= 0 .
\end{align}
\end{subequations}

Solving the system \eqref{eq:sys}, we express the fluctuations of the hydrodynamic velocity and electron density in terms of 
$\mathcal{F}_{\bm q,\omega}$ and $\bm q\cdot\bm{\mathcal{J}}_{-\bm q,-\omega}$ and use Eq.~\eqref{eq:noise_q} to determine the fluctuation contribution to the transport current in Eq.~\eqref{eq:current_average}. 

To be specific, we assume that a uniform external electric field $\bm{E}$ is applied in the $x$-direction (along the length $L$ of the Hall bar) and express the macroscopic conductivity of the sample in the form $\sigma = \sigma_0 + \sigma_{\mathrm{fl}}$.
A straightforward  calculation outlined in Appendix \ref{sec:derivation} gives the following result for the fluctuation contribution to the conductivity, $\sigma_{\mathrm{fl}}$:
\begin{align}
    \label{eq:master}
    \sigma_{\rm fl}
     & \, =  \frac{T\sigma_0}{\rho}\, \int \frac{d^2 q}{(2\pi)^2}\,  \frac{q^2_{y}}{\gamma_\sigma\left(\gamma_\sigma + \gamma_\nu + \gamma_p + \gamma_H\right)}.
\end{align}

Below, we assume that the Coulomb interaction is screened by a gate located at a distance $d$ from the sample. Therefore, for wave vectors $q \lesssim 1/d$, the interaction becomes local, $U(q) \approx 4 \pi e^2 d/ \varepsilon$, and the charge density fluctuations relax by diffusion,  $\gamma_\sigma \propto q^2$. In this regime, the fluctuation contribution to the conductivity 
may be expressed in the form
\begin{align}
    \label{eq:sigma_fl_q_H}
    \sigma_{\rm fl}
     = & \, \frac{T \sigma_{0}}{\rho \nu^2} \frac{\alpha^2}{1 + \alpha} \int \frac{dq_{x} dq_{y}}{(2\pi)^2} \frac{q^2_{y}}{q^2 \left(q^2 + \ell^{-2}_{G} + \ell^{-2}_{H}\right)}.
\end{align}
Here $\alpha$ is the dimensionless ratio of the viscous and Maxwell relaxation rates,
\begin{align}
\label{eq:alpha}
 \alpha^{-1} \equiv \frac{\gamma_\sigma}{\gamma_\nu} = \frac{\sigma_0}{e^2\nu}\, \left[ (\partial_\mu n)^{-1} + \frac{4\pi e^2 d}{\varepsilon} \right],
\end{align}
and $\ell_{G}$  is the Gurzhi length~\cite{Gurzhi1963} associated with momentum-relaxing scattering and  $\ell_{H}$ is the characteristic length scale associated with the magnetic field,  
\begin{subequations}
        \label{eq:l_d_l_H_def}
\begin{align}
    \label{eq:l_d_def}
    \frac{1}{\ell_{G}} = & \,  \sqrt{\frac{\gamma_p}{  \nu \left( 1+ \alpha^{-1}\right)}} = \sqrt{\frac{k}{  \eta \left( 1+ \alpha^{-1}\right)}}, \\
    \frac{1}{\ell_{H}} = & \, \sqrt{ \frac{ \gamma_H}{  \nu \left( 1+ \alpha^{-1}\right)}} = \frac{H}{c}  \sqrt{\frac{ \sigma_0}{  \eta \left( 1+ \alpha^{-1}\right)}}.
      \label{eq:l_H_def}
\end{align}
\end{subequations}
The magnetic length scale $\ell_H$ corresponds to momentum relaxation caused by magnetic friction. In monolayer graphene (MLG), $\alpha$ is on the order of unity, and $\ell_H$ may be estimated as  $\ell_H  \sim l_H^2/ l_{\mathrm{ee}} \sim l_H^2/\lambda_T$.

The logarithmic divergence of the
integral in Eq.~\eqref{eq:sigma_fl_q_H} at large momenta is cut off at the microscopic scale  $q_0 \sim 2\pi/\mathrm{max} \left\{d, l_{\mathrm{ee}} \right\}$. In bulk samples, the logarithmic divergence at small momenta is cut off at the momentum scale $\sqrt{\ell^{-2}_{G} + \ell^{-2}_{H}}$ associated with momentum relaxation caused by disorder and magnetic friction. Under our assumptions $\ell^{-2}_{G} + \ell^{-2}_{H} \ll q_0^2$. Therefore, at weak magnetic fields magnetoconductivity is dominated by fluctuations in a wide interval momenta, $\sqrt{\ell^{-2}_{G} + \ell^{-2}_{H}} \lesssim q \lesssim  q_0$. Performing a straightforward integration in Eq.~\eqref{eq:sigma_fl_q_H}, gives
the fluctuation conductivity of bulk samples,
\begin{align}
    \label{eq:result_bulk}
    \frac{\sigma_{\mathrm{fl}}}{\sigma_{0}} = & \, \frac{ \varsigma}{2\pi}\, \ln \frac{q_0^2}{\ell^{-2}_{G} + \ell^{-2}_{H}}.
\end{align}
Here we introduced a dimensionless parameter 
\begin{align}
\label{eq:varsigma_def}
\varsigma = & \,  \frac{T\alpha}{4\rho\nu^2 \left(1+\alpha^{-1}\right)},
\end{align}
which depends on the properties of the electron liquid and characterizes the magnitude of the fluctuation conductivity. For a strongly interacting electron liquid in monolayer graphene $\varsigma$ is of order unity. Therefore, strong magnetoresistance arises at characteristic magnetic field scale $H_{\mathrm{dis}}$ at which $\ell_H = \ell_G$;
\begin{align}
    \label{eq:H_disorder}
    H_{\mathrm{dis}} = & \, c \sqrt{ \frac{k}{\sigma_0}}  
    = \frac{c}{\ell_G} \sqrt{\frac{\eta (1+\alpha^{-1})}{\sigma_0}}.
\end{align}
At $H\sim H_{\mathrm{dis}}$ the magnetic length may be estimated as 
$l_H\sim \sqrt{l_{\mathrm{ee}} \ell_G} \gg l_{\mathrm{ee}} $. Therefore, in the hydrodynamic regime, strong MR appears at weak fields, at which the influence of the magnetic field on properties of the electron liquid may be neglected. For long-range puddle disorder, the friction coefficient is related to the variance of the disorder-induced electron density modulation 
$\left\langle\left(\delta n\right)^2\right\rangle$ by~\cite{Li2020}
\begin{align}
    \label{eq:k_puddle}
    k = \frac{e^2}{2\sigma_0}\left\langle\left(\delta n\right)^2\right\rangle .
\end{align}
for $\delta n\approx 2\times10^{9}\,\mathrm{cm}^{-2}$ and $\sigma_0\sim 4e^2/h$  Eq.~\eqref{eq:H_disorder} gives a relatively weak field scale
\begin{align}
    \label{eq:Hdis_estimate}
    H_{\mathrm{dis}} = \frac{c\,e}{\sigma_0}\sqrt{\frac{\langle(\delta n)^2\rangle}{2}} \approx 15\ \mathrm{mT},
\end{align}
which increases with disorder strength.

Equation Eq.~\eqref{eq:result_bulk} also describes the fluctuation contribution conductivity of finite samples, provided their smallest dimension exceeds either the Gurzhi length $\ell_G$ or the magnetic length scale $\ell_H$. In clean graphene devices, the sample size may exceed the Gurzhi length. In this case, at weak magnetic fields, the infrared divergence of the momentum integral in Eq.~\eqref{eq:sigma_fl_q_H} is cut off by the sample size.  For a Hall bar of length $L$ and width $W$, the integral over $q$ in Eq.~\eqref{eq:sigma_fl_q_H} may be evaluated analytically for an arbitrary aspect ratio of the sample $L/W$. The rather cumbersome result, Eq.~\eqref{eq:Main_formula_x_y_z}, is presented, along with the derivation in Appendix \ref{sec:general_aspect_ratio}. In the main text, we focus on the special cases of square ($W=L$), wide ($W\gg L$) and long ($W\ll L$) samples, for which the result takes a considerably simpler form. In the equations below $\ell = \max\{l_{\rm ee},\, d\} $ is the short distance cutoff.

For long samples, $L \gg W$, we obtain
\begin{align}
    \label{eq:barresult_long_q_h}
     \frac{\sigma_{\rm fl}}{\sigma_0} & \, = \frac{\varsigma}{\pi} \left[\ln \left(\frac{W}{\ell}\right) - \ln \left(1 + \sqrt{1 + \frac{W^2 \left(\ell^{-2}_{G} + \ell^{-2}_{H}\right)}{\left(2 \pi\right)^2}}\right) \right. \nonumber \\
    + & \, \left. \frac{\left(2 \pi\right)^2}{W^2 \left(\ell^{-2}_{G} + \ell^{-2}_{H}\right)} \left(\sqrt{1 + \frac{W^2 \left(\ell^{-2}_{G} + \ell^{-2}_{H}\right)}{\left(2 \pi\right)^2}} - 1\right)\right].
\end{align}
For square samples, $W = L$, we find 
\begin{align}
    \label{eq:result_square_q_H}
    \frac{\sigma_{\rm fl}}{\sigma_0} = & \, \frac{\varsigma}{\pi} \left[\ln \left(\frac{L}{\ell}\right) - \ln \left(1 + \sqrt{1 + \frac{L^2 \left(\ell^{-2}_{G} + \ell^{-2}_{H}\right)}{\left(2 \pi\right)^2}}\right)\right].
\end{align}
For wide Hall bars, $W\gg L$, we get 
\begin{align}
    \label{eq:barresult_wide_q_h}
    \frac{\sigma_{\rm fl}}{\sigma_0} = &\, \frac{\varsigma}{\pi} \left[\ln \left(\frac{L}{\ell}\right) - \ln \left(1 + \sqrt{1 + \frac{L^2 \left(\ell^{-2}_{G} + \ell^{-2}_{H}\right)}{\left(2 \pi\right)^2}}\right)  \right. \nonumber \\
    & \left. - \frac{\left(2 \pi\right)^2}{L^2 \left(\ell^{-2}_{G} + \ell^{-2}_{H}\right)} \left(\sqrt{1 + \frac{L^2 \left(\ell^{-2}_{G} + \ell^{-2}_{H}\right)}{\left(2 \pi\right)^2}} - 1\right)\right].
\end{align}

\begin{figure}[t]
    \centering
    \includegraphics[width=\columnwidth]{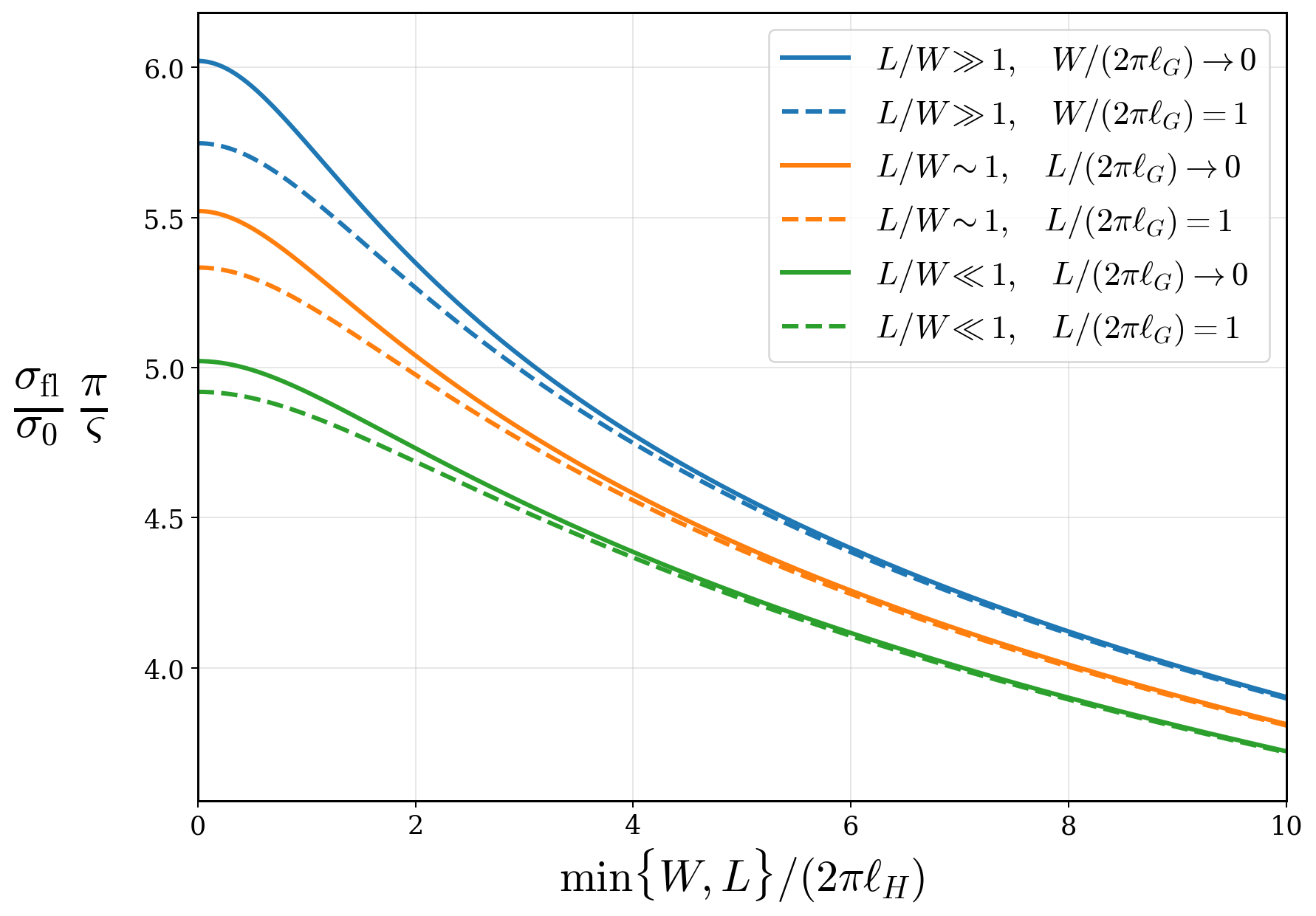}
    \caption{Fluctuation conductivity of long (blue), square (orange), and wide (green) samples, given by Eqs. \eqref{eq:barresult_long_q_h}, \eqref{eq:result_square_q_H}, and \eqref{eq:barresult_wide_q_h}
   is plotted as a function of $H/H_{\mathrm{size}} = \mathrm{min}\{W,L\}/(2\pi \ell_H)$. The solid lines correspond to pristine samples, $\mathrm{min} \{W, L \}/2 \pi \ell_G \to 0$, and the dashed lines correspond to weakly disordered samples, $\mathrm{min} \{W, L \}/2 \pi \ell_G = 1$. In the plots we set
      $\frac{W}{\ell} = 500, \, \frac{L}{\ell} = 500$ .}
    \label{fig:MR}
\end{figure}

In Eqs.~\eqref{eq:barresult_long_q_h}--\eqref{eq:barresult_wide_q_h} the magnetic field and disorder strength enter only through the combination $\mathrm{min}\{L^2,W^2\}\,(\ell_G^{-2}+\ell_H^{-2})/(2\pi)^2$, which by Eqs.~\eqref{eq:l_d_l_H_def} and \eqref{eq:H_disorder} may be written as
\begin{align}
    \label{eq:field_variable}
    \frac{\mathrm{min}\{L^2,W^2\}}{(2\pi)^2}\left(\frac{1}{\ell_G^{2}}+\frac{1}{\ell_H^{2}}\right) = \frac{H^2 + H_{\mathrm{dis}}^2}{H_{\mathrm{size}}^2},
\end{align}
where
\begin{align}
    \label{eq:H_size}
    H_{\mathrm{size}} = \frac{2\pi c}{\mathrm{min}\{L,W\}}\sqrt{\frac{\eta\left(1+\alpha^{-1}\right)}{\sigma_0}}
\end{align}
is the field at which the magnetic length scale $\ell_H$ becomes comparable to the smaller dimension of the Hall bar. The characteristic field $H_{\mathrm{size}}$ corresponds to classically weak fields. For a strongly interacting Dirac liquid in monolayer graphene $\alpha\sim1$, $\nu\sim v\lambda_T$, and $\eta = \rho\nu$ with $\rho\sim T^3/v^4$, so that the magnetic length scale in Eq.~\eqref{eq:l_H_def} may be estimated as $\ell_H\sim l_H^2/\lambda_T$, where $l_H = \sqrt{\hbar c/|e|H}$ is the magnetic length. The condition $\ell_H\sim \mathrm{min}\{L,W\}/2\pi$ is therefore reached at $l_H\sim\sqrt{\lambda_T\,\mathrm{min}\{L,W\}}\gg\lambda_T$, where the dependence of the properties of the electron liquid on the magnetic field is negligible, as assumed in Sec.~\ref{sec:hydro}. For Hall bars a few $\mu$m wide and $\lambda_T \approx 80\,\mathrm{nm}\times(100\,\mathrm{K}/T)$, Eq.~\eqref{eq:H_size} gives $H_{\mathrm{size}}$ of a few tens of millitesla, growing with temperature.

In relatively clean samples, $\ell_G \gtrsim \mathrm{min}\{L,W\}$, the fluctuation magnetoresistance described by Eqs.~\eqref{eq:barresult_long_q_h}, \eqref{eq:result_square_q_H}, and \eqref{eq:barresult_wide_q_h} is controlled by the finite size effects at weak magnetic fields, $H\lesssim H_{\mathrm{size}}$. The functional form of the weak-field MR has a pronounced dependence on the aspect ratio $W/L$ of the Hall bar shown in Fig~\ref{fig:MR}. Disorder-induced momentum relaxation leads to rounding of MR curves at weak fields, as shown by dashed lines. At stronger fields, $H \gg H_{\mathrm{size}}$,  the fluctuation conductivity of finite samples approaches the bulk conductivity given by Eq.~\eqref{eq:result_bulk}.

\section{Discussion of the results\label{sec:summary}}

We developed a theory of magnetoresistance in graphene Hall bars at charge neutrality, which is caused by the fluctuation mechanism introduced in Ref.~\cite{fluct_graphene}. This mechanism describes MR of relatively clean graphene systems at weak magnetic fields, where the magnetic length $l_H$ exceeds the microscopic scale corresponding to the onset of the hydrodynamic regime. In this regime, the intrinsic conductivity $\sigma_0$ and other parameters of the electron liquid may be assumed independent of $H$. Nevertheless, the macroscopic conductivity of the system exhibits strong dependence on $H$, which arises from the coupling of the charge density fluctuations and the fluctuations of the hydrodynamic flow.

The consideration of Ref.~\cite{fluct_graphene} focused on square samples and was restricted to the fluctuations driven by the Nyquist-Johnson noise associated with the intrinsic conductivity of the electron liquid. Neglecting other sources of fluctuations was based on the observation that at $H=0$ the fluctuations caused by the random viscous stresses are decoupled from the charge density fluctuations. Here we showed that at $H\neq 0$, the random viscous stresses couple to the charge density fluctuations and produce a contribution to MR, which is comparable with the contribution caused by the Nyquist-Johnson noise. We also included the effect of weak momentum on MR. The fluctuation magnetoconductivity of bulk samples,   Eq.~\eqref{eq:result_bulk}, is controlled by the Gurzhi length. For finite size Hall bars the fluctuation magnetoconductivity depends on 
both the length and the width of the bar and is described by Eqs.~\eqref{eq:barresult_long_q_h} - \eqref{eq:barresult_wide_q_h}
and also Eq.~\eqref{eq:Main_formula_x_y_z} in Appendix \ref{sec:general_aspect_ratio}.

The functional form of the fluctuation-induced magnetoconductivity $\sigma_{\mathrm{fl}} (H)$, depends on the aspect ratio of the Hall bar, $W/L$, and also on disorder strength, as illustrated in Fig.~\ref{fig:MR}. 
Another distinct feature of the fluctuation mechanism is that strong MR arises at relatively weak magnetic fields characteristic magnetic fields that depend on the smaller dimension of the sample, $H_{\mathrm{size}}$ in Eq.~\eqref{eq:H_size}, and disorder strength, $H_{\mathrm{dis}}$ in Eq.~\eqref{eq:H_disorder}. 
This, and the dependence of magnetoconductivity on the Hall bar geometry and its orientation relative to the current, may help distinguish the fluctuation mechanism of MR from other mechanisms in experiments.

Another mechanism of MR at charge neutrality was introduced in Ref.~\onlinecite{MullerSachdev2008}  (see also \cite{Levchenko2024}). It is associated with the hydrodynamic flow driven by the Lorentz force. In Hall bars this mechanism is inactive because the macroscopic flow is precluded by the boundary conditions.

Although we focused our discussion on graphene samples, our results may also apply to other clean two-dimensional semimetals.
The importance of fluctuations in hydrodynamics of low-dimensional systems has been widely appreciated~\cite{ForsterNelsonStephen1977,AndreevJETP1980}. From this perspective, our work extends the theory of hydrodynamic fluctuations in liquids with Galilean invariance~\cite{LandauLifshitz1957} to electron liquids at charge neutrality. 

\begin{acknowledgments} We are especially grateful to Matt Yankowitz for numerous stimulating discussions that initiated this work. We also benefited from discussions with Boris Spivak and David Cobden. The work of A. L. was supported by NSF Grant No. DMR-2452658 and H. I. Romnes Faculty Fellowship provided by the University of Wisconsin-Madison Office of the Vice Chancellor for Research and Graduate Education with funding from the Wisconsin Alumni Research Foundation. The work of A.G. and A. V. A. was supported by the National Science Foundation (NSF) Grant No. DMR-2424364.
\end{acknowledgments}

\appendix
 
\section{Derivation of Eq.~\eqref{eq:master} \label{sec:derivation}}

In this Appendix, we derive Eq.~\eqref{eq:master} for the fluctuation contribution to the macroscopic conductivity of the system starting from  Eq.~\eqref{eq:sys}.

Solving the linear  system \eqref{eq:sys}, we express the Fourier amplitudes of the fluctuations of density and transverse velocity in terms of the fluctuating amplitudes of the Langevin sources:
\begin{subequations}
    \label{eq:n_u_sol}
\begin{align}
    \label{eq:nsol}
      e n_{\bm{q},\omega} = & \,    \frac{- i  \left(- i \omega  + \gamma_p +\gamma_\nu \right) \bm{q}\cdot \bm{\mathcal{J}}_{\bm{q},\omega} - \frac{\sigma_0}{c\rho}\,\mathcal{F}_{\bm q,\omega}}{\mathcal{D}(\bm q,\omega)} ,\\
      \label{eq:usol}
    \bm{u}_{\bm{q},\omega} = & \,\frac{ - i \, \hat{z}\times \bm{q}}{\rho\, q^2  \mathcal{D}(\bm q,\omega)} \Bigg\{ \left[\bm{E}\cdot[\hat z\times\bm q] - \omega\, \frac{H}{c}\right]\bm{q}\cdot \bm{\mathcal{J}}_{\bm{q},\omega} \nonumber\\
    & \, -i \frac{\omega+ i \gamma_\sigma}{H}\mathcal{F}_{\bm q,\omega} \Bigg\}.
\end{align}
\end{subequations}
Here we introduced a shorthand notation
\begin{align}
            \label{eq:D_def}
    \mathcal{D}(\bm q,\omega) = & \left( - i \omega  + \gamma_{\sigma} \right)\left(- i  \omega + \gamma_p + \gamma_\nu +\gamma_H\right) \nonumber \\
     & + i\, \frac{\sigma_0\, \bm{q} \cdot [  \bm{E} \times \bm{H}]}{c \rho} - \gamma_{H}\, \gamma_{\sigma}.
\end{align}    
The fluctuation contribution to the transport current, which is given by the second term on the right-hand side of Eq.~\eqref{eq:current_average}, is expressed in terms of the Fourier amplitudes as follows,
\begin{align}
\label{eq:jfl_def}
\bm j_{\rm fl} = \int\!\frac{d\omega\, d^2q}{(2\pi)^3}\,\big\langle e n_{-\bm q,-\omega}\, \bm u_{\bm q,\omega}\big\rangle .
\end{align}

Using Eqs.~\eqref{eq:n_u_sol} and \eqref{eq:noise_q}, and noting that
$\mathcal D(-\bm q,-\omega) = \mathcal D^*(\bm q,\omega)$, 
we express the average in the integrand of Eq.~\eqref{eq:current_average} in the form
\begin{align}
\label{eq:product}
\big\langle e n_{-\bm q,-\omega}\, \bm u_{\bm q,\omega}\big\rangle
= \frac{2T\sigma_0 q^2}{\rho}\;\frac{\hat z\times\bm q}{q^2}\;\frac{\mathcal N(\bm q,\omega)}{|\mathcal{D}(\bm q,\omega)|^2},
\end{align}
where  
\begin{align}
\label{eq:D_squared}
|\mathcal{D}(\bm q,\omega)|^2 = & \,  \left[\omega^2 -\gamma_\sigma(\gamma_\nu + \gamma_p)\right]^2 \nonumber \\
+ & \, \left[\omega(\gamma_\sigma + \gamma_\nu + \gamma_p + \gamma_H) - \frac{\sigma_0}{c\rho}\bm q\cdot[\bm E\times\bm H]\right]^2.
\end{align}
and we introduced a shorthand notation
\begin{align}
\label{eq:N_final}
\mathcal N = & \,  \big\{i \omega + \gamma_\nu + \gamma_p\big\} \bm{E} \cdot \big[\hat{z} \times \bm{q}\big] \nonumber \\
- & \, i \frac{H}{c} \big\{\omega^2 + \gamma_{\sigma}\left(\gamma_\nu + \gamma_p\right)\big\}.
\end{align}
The real Hall-type terms $\mp \left(\gamma_p + \gamma_\nu\right) \omega H/c$ have canceled. This cancellation is the central structural feature of the theory, and Eq.~\eqref{eq:N_final} makes its origin explicit: the term generated by the current noise [the $i\omega H/c$ part of the advection numerator in Eq.~\eqref{eq:usol}, beating against the damping factor $(i\omega+\gamma_p + \gamma_\nu)$ of the density response] is removed, exactly and pointwise in $(\bm q,\omega)$, by the force-noise contribution. The two terms share the same denominator $|\mathcal D|^2$, so no expansion in the applied field $\bm E$ is involved: the cancellation is a generic property of the FDT-consistent noise set and holds to all orders in the drive. Its magnitude is fixed by the FDT itself -- the force-noise weight $2T\rho \left(\gamma_\nu + \gamma_p\right) H^2 q^2$ is precisely $2T$ times the (viscous plus disorder) damping of the transverse channel, and any other value would leave a remainder. The physical necessity of the cancellation is seen at $\bm E = 0$: the real part of Eq.~\eqref{eq:N_final} then vanishes identically, so no equilibrium current is generated, and the same noise balance guarantees the equal-time equipartition $\rho\langle|u^{(t)}|^2\rangle_{\bm q} = T$ mode by mode. Had the force noises been omitted, the surviving interference term would have violated both properties and distorted the shape of the magnetoresistance crossover.

The imaginary part of Eq.~\eqref{eq:N_final} does not contribute to Eq.~\eqref{eq:jfl_def}: under $(\bm q,\omega)\to(-\bm q,-\omega)$ the factor $|\mathcal D|^2$ is invariant while the prefactor $\hat z\times\bm q$ is odd and both imaginary terms are even, so they multiply an odd integrand and vanish upon integration, as they must for $\bm j_{\rm fl}$ to be real. The surviving real part is already linear in $\bm E$; to linear response, the denominator may therefore be evaluated at zero field, $|\mathcal D|^2 \to |\mathcal{D}_{0}|^2$, with 
\begin{align}
\label{eq:D_0_def}
|\mathcal{D}_{0} (\bm q,\omega)|^2 = & \,  \omega^2\left( \gamma_\sigma+\gamma_p+\gamma_\nu+\gamma_H \right)^2 \nonumber \\
&+ \left(\omega^2-\gamma_\sigma \left[ \gamma_p + \gamma_\nu\right] \right)^2.
\end{align}

To linear order in the electric field (only keeping even in $\omega$ terms)
\begin{align}
& \, \big\langle e n_{-\bm q,-\omega} \, \bm u_{\bm q,\omega} \big \rangle =
\frac{2T \sigma_0 q^2}{\rho} \frac{\hat{\bm z}\times\bm q}{q^2} \,
\bm E\cdot[\hat{\bm z}\times\bm q]
\nonumber \\
\times & \, \frac{\gamma_p + \gamma_{\nu}}{|\mathcal{D}_{0} (\bm q,\omega)|^2}.
\end{align}
Therefore, for the current we arrive at 
\begin{align}
    \label{eq:jfl_result}
    \bm j_{\rm fl}
     =  \int \frac{d \omega\, d^2 q}{(2\pi)^3}\, \frac{2T\sigma_0 q^2}{\rho}\;  \frac{(\hat{z}\times \bm{q})\left( \bm{E}\cdot [   \hat{z} \times \bm{q}]\right)}{q^2} \frac{\gamma_p + \gamma_{\nu}}{|\mathcal{D}_{0} (\bm q,\omega)|^2}.
\end{align}
For our geometry, in which the bias electric field $\bm{E}$ is directed along the $x$-axis, we obtain the fluctuation contribution to the macroscopic conductivity in the form
\begin{align}
    \label{eq:jfl_result}
    \sigma_{\rm fl}
     =  \frac{2T\sigma_0 }{\rho}\int \frac{d \omega\, d^2 q}{(2\pi)^3}\,   \frac{ q^2_{y} \left(\gamma_p + \gamma_{\nu}\right)}{|\mathcal{D}_{0} (\bm q,\omega)|^2}.
\end{align}
Substituting Eq.~\eqref{eq:D_0_def} in the above equation and integrating over $\omega$ we obtain Eq.~\eqref{eq:master}.

\section{Estimation of the longitudinal contribution to MR\label{sec:longitudinal}}
In this Appendix we show that the longitudinal contribution to magnetoresistance is reduced by $\left(\gamma_p \tau_{\mathrm{ee}}\right)^2$ at semiclassically weak magnetic fields compared to the transverse contribution, and therefore can be neglected. We use the same Langevin description of thermal fluctuations of the electron liquid in the hydrodynamic regime at the charge neutrality point as in the main text. 

The longitudinal component of the linearized Navier-Stokes-Langevin equation has the form 
\begin{align}
    \label{eq:u_evolution_l}
    \rho\,\partial_t \bm{u}^{(l)} = & \,  e n \bm{E}^{(l)} + \frac{1}{c}\,[\bm{j}^{(t)}\times \bm{H}]
    \nonumber \\
    & - k \,\bm u^{(l)}+ \left(\eta + \zeta\right) \nabla^2 \bm{u}^{(l)}  + \bm f^{(l)}  - \bm{\nabla}P.
\end{align}
Here, the pressure is related to the entropy density $s$ by 
\begin{align}
\label{eq:P_s}
    \bm{\nabla} P = & \, \left(\partial_s P\right)_n \bm{\nabla} s.
\end{align}
For linear fluctuations, the entropy evolution equation reduces to a continuity equation,
\begin{align}
    \label{eq:entropy_continuity}
    \partial_t s + \bm{\nabla} \cdot \bm{j}_s= & \, 0.
\end{align}
Here the entropy current is given by 
\begin{align}
    \label{eq:j_s}
    \bm{j}_s = & \, s_0 \bm{u} - D_s \bm{\nabla} s + \bm{\mathcal{J}}^s,
\end{align}
where $\bm{\mathcal{J}}^s$ is a random Langevin source.

Substituting \eqref{eq:j_s} into \eqref{eq:entropy_continuity} we get in Fourier components,
\begin{align}
    \label{eq:s_q}
   - i \omega  s_{\bm{q},\omega} = & \,  - i s_0 \bm{q} \cdot \bm{u}^{(l)}_{\bm{q},\omega} - Dq^2 s_{\bm{q},\omega} - i \bm{q}\cdot \bm{\mathcal{J}}^s_{\bm{q},\omega}, \nonumber \\
   s_{\bm{q},\omega} = & \, \frac{  \bm{q}\cdot 
   \left( s_0 \bm{u}^{(l)}_{\bm{q},\omega} + \bm{\mathcal{J}}^s_{\bm{q},\omega} \right)  }{\omega + i D_s q^2} 
\end{align}
Taking the time-derivative of Eq.~\eqref{eq:u_evolution_l} we get 
\begin{align}
    \label{eq:long_fluct}
    - & \,  \rho \omega^2 \bm{u}^{(l)}_{\bm{q},\omega} =  - i \omega e n_{\bm{q},\omega}\bm{E}^{(l)} - i  \frac{\omega}{c}\,[\bm{j}^{(t)}_{\bm{q},\omega}\times \bm{H}]
    \nonumber \\
    & + i \omega  k \,\bm u^{(l)}_{\bm{q},\omega} + i \omega \left(\eta + \zeta\right) q^2 \bm{u}^{(l)}_{\bm{q},\omega}  - i \omega \bm f^{(l)}_{\bm{q},\omega}  - \omega \bm{q} P_{\bm{q},\omega}.
\end{align}

Using \eqref{eq:s_q} and \eqref{eq:P_s} we obtain
\begin{align}
 \label{eq:pressure_gradient_U}
   - \omega \bm{q} P_{\bm{q},\omega} = & \, - \left( \partial_s P\right)_n  \omega \bm{q} s_{\bm{q},\omega}  \nonumber \\
   = & \, - \left( \partial_s P\right)_n   s_0 \, \frac{  \omega q^2 
    }{\omega + i D_s q^2} \, \left( \bm{u}^{(l)}_{\bm{q},\omega} + \frac{1}{ s_0}\bm{\mathcal{J}}^{s,(l)}_{\bm{q},\omega} \right).  
\end{align}
Substituting this into Eq.~\eqref{eq:long_fluct} we get 
\begin{align}
    \label{eq:u_l_resolvent}
  & \, \left[ \omega q^2 \frac{\left( \partial_s P\right)_n   s_0  
    }{\omega + i D_s q^2}   - \rho \omega^2 
    -  i \omega  k  - i  \left(\eta + \zeta\right) \omega q^2  \right] \bm{u}^{(l)}_{\bm{q},\omega} \nonumber \\
    = & \,  - i \omega e n_{\bm{q},\omega}\bm{E}^{(l)} - i  \frac{\omega}{c}\,[\bm{j}^{(t)}_{\bm{q},\omega}\times \bm{H}]
    \nonumber \\
    & \, - i \omega \bm f^{(l)}_{\bm{q},\omega} 
    -   \omega q^2 \, \frac{\left(\partial_s P\right)_n}{\omega + i D_s q^2} \,  \bm{\mathcal{J}}^{s,(l)}_{\bm{q},\omega} .
\end{align}
Using Eqs. \eqref{eq:current} and \eqref{eq:EMF_def}, we obtain the transverse Fourier component of the current as follows
\begin{align}
\label{eq:Current_transv_comp_Fourier}
    \bm{j}^{(t)}_{\bm{q},\omega}
    = & \, \sigma_0 \bm{E}^{(t)}_{\bm{q},\omega}
    + \frac{\sigma_0}{c}
    \left[\bm{u}^{(l)}_{\bm{q},\omega} \times \bm{H}\right]
    + \bm{\mathcal{J}}^{(t)}_{\bm{q},\omega}.
\end{align}
Substituting \eqref{eq:Current_transv_comp_Fourier} in Eq. \eqref{eq:u_l_resolvent} we obtain
\begin{align}
& \, \left[-q^2 \frac{\left(\partial_s P\right)_n s_0}{\omega + i D_s q^2}
+ \rho \omega + i k + i (\eta + \zeta) q^2
+ \frac{i \sigma_0 H^2}{c^2}\right] \bm{u}^{(l)}_{\bm{q},\omega}
\nonumber \\
= & \, i e n_{\bm{q},\omega} \bm{E}^{(l)} + \frac{i}{c} \left[\bm{\mathcal{J}}^{(t)}_{\bm{q},\omega}\times\bm{H}
\right] + i \bm{f}^{(l)}_{\bm{q},\omega} + q^2
\frac{\left(\partial_s P\right)_n}{\omega + i D_s q^2} \bm{\mathcal{J}}^{s,(l)}_{\bm{q},\omega}.
\end{align}
The Langevin terms on the right-hand side can be omitted when evaluating the microscopic current, since the corresponding sources are uncorrelated with those entering the density fluctuation in Eq. \eqref{eq:nsol}. Their mixed correlators therefore vanish upon averaging. Consequently, the longitudinal velocity component relevant for the current is given by
\begin{align}
\label{eq:u_l_result}
\bm{u}^{(l)}_{\bm{q},\omega} = & \, 
\frac{ i e n_{\bm{q},\omega} \left(\omega + i \gamma_s\right) \bm{E}^{(l)}}{ \rho \left[\omega (\omega + i \gamma_s) - c^2_{2} q^2  + i (\omega + i \gamma_s)\left(\gamma_{k} + \gamma_{H} + \gamma_{\zeta}\right)\right]},
\end{align}
where $c^2_{2} = \frac{\left(\partial_s P\right)_n s_0}{\rho}$, \, $\gamma_s=D_sq^2$, and $\gamma_{k}, \, \gamma_{H}, \, \gamma_{\zeta}$ are given by Eq. \eqref{eq:rates}.
Defining 
\begin{align}
    \label{eq:Gamma_def}
    \Gamma_s = & \, \gamma_{k} + \gamma_{\zeta} + \gamma_{s}, \quad \Gamma = \gamma_k+ \gamma_\nu + \gamma_\sigma, \quad \Sigma = \Gamma + \gamma_H.
\end{align}
we write
\begin{widetext}
    \begin{align}
    \label{eq:u_l_Gamma}
    & \, \bm{u}^{(l)}_{\bm{q},\omega} = \frac{i e n_{\bm{q},\omega} \bm{E}^{(l)}}{\rho} \frac{\omega + i \gamma_s}{\left[\left(\omega + \frac{i}{2} \left(\Gamma_{s} + \gamma_{H}\right) \right)^2  - c^2_{2} q^2 
    + \frac{1}{4} \left(\Gamma_{s} + \gamma_{H}\right)^2 - \gamma_s\left(\Gamma_{s} + \gamma_{H} -\gamma_s  \right)\right]}.
\end{align}
Both poles are in the LHP.

The longitudinal contribution to the fluctuation conductivity is 
\begin{align}
\label{eq:longitudinal_conductivity}
    \sigma^{(l)}_{\mathrm{fl}} = & \, \frac{i}{\rho} \int \frac{d \omega}{2\pi} (dq) \frac{q_x^2}{q^2} \frac{ e^2  \langle |n_{\bm{q},\omega}|^2 \rangle\left(\omega + i \gamma_s \right)}{\left[\left(\omega + \frac{i}{2} \left(\Gamma_{s} + \gamma_{H}\right) \right)^2  - c^2_{2} q^2 + \left(\frac{\Gamma_{s}}{2} + \frac{\gamma_{H}}{2} - \gamma_{s}\right)^2\right]}.
\end{align}
\end{widetext}
Using Eq.~\eqref{eq:nsol} and Eq.~\eqref{eq:noise_q}, we get 
\begin{align}
      & \, e^2  \langle |n_{\bm{q},\omega}|^2 \rangle = 2 T \sigma_0 q^2
 \frac{ (\omega^2 + \left[\gamma_{\nu} + \gamma_{p}\right]^2) + \left[\gamma_{\nu} + \gamma_{p}\right] \, \gamma_H}{|\mathcal{D}_0(\bm q,\omega)|^2}.
\end{align} 
Here
\begin{align}
    \label{eq:D_0_def}
    \mathcal{D}_0 (\bm q,\omega) = & \, \left( - i \omega  + \gamma_{\sigma} \right)\left(- i  \omega + \gamma_{\nu} + \gamma_{p} +\gamma_H\right)  - \gamma_{H}\, \gamma_{\sigma}.
\end{align}
The integration contour over $\omega$ in Eq.~\eqref{eq:longitudinal_conductivity} should be closed in the UHP and is determined by the residues of $\bm{D}_0$.
\begin{widetext}
\begin{align}
\label{eq:longitudinal_conductivity}
    \sigma^{(l)}_{\mathrm{fl}} = & \, i \frac{2 T \sigma_0}{\rho} \int  (dq) q_x^2  \int \frac{d \omega}{2\pi} \frac{(\omega^2 + \left[\gamma_{\nu} + \gamma_{p}\right]^2) + \left[\gamma_{\nu} + \gamma_{p}\right] \, \gamma_H}{\left(\omega^2 -\gamma_{\sigma} \left[\gamma_{\nu} + \gamma_{p}\right] \right)^2 + \omega^2 \left(\Gamma + \gamma_{H}\right)^2} \frac{\left(\omega + i \gamma_s \right)}{\left[\left(\omega + \frac{i}{2} \left(\Gamma_{s} + \gamma_{H}\right) \right)^2  - c^2_{2} q^2 + \left(\frac{\Gamma_{s}}{2} + \frac{\gamma_{H}}{2} - \gamma_{s}\right)^2\right]},
\end{align}
From here, the magnetoconductivity $\frac{\delta \sigma^{(l)}_{\mathrm{fl}}}{\sigma_0} = \frac{\sigma^{(l)}_{\mathrm{fl}} \left(H\right) - \sigma^{(l)}_{\mathrm{fl}} \left(H = 0\right)}{\sigma_{0}}$ is given by 
    \begin{align}
    & \, \frac{\delta \sigma^{(l)}_{\mathrm{fl}}}{\sigma_0} = i \frac{2 T}{\rho} \int  (dq) q_x^2  \int \frac{d \omega}{2\pi} \frac{\left(\omega + i \gamma_s \right)}{\left[\left(\omega + \frac{i}{2} \left(\Gamma_{s} + \gamma_{H}\right) \right)^2  - c^2_{2} q^2 + \left(\frac{\Gamma_{s}}{2} + \frac{\gamma_{H}}{2} - \gamma_{s}\right)^2\right]} \frac{(\omega^2 + \left[\gamma_{\nu} + \gamma_{p}\right]^2) + \left[\gamma_{\nu} + \gamma_{p}\right] \, \gamma_H}{\left(\omega^2 -\gamma_{\sigma} \left[\gamma_{\nu} + \gamma_{p}\right]\right)^2 + \omega^2 \left(\Gamma + \gamma_{H}\right)^2} \nonumber \\
    - & \, i \frac{2 T}{\rho} \int  (dq) q_x^2  \int \frac{d \omega}{2\pi} \frac{\left(\omega + i \gamma_s \right)}{\left[\left(\omega + \frac{i}{2} \Gamma_{s}\right)^2  - c^2_{2} q^2 + \left(\frac{\Gamma_{s}}{2} - \gamma_{s}\right)^2\right]} \frac{\omega^2 + \left[\gamma_{\nu} + \gamma_{p}\right]^2}{\left(\omega^2 -\gamma_{\sigma} \left[\gamma_{\nu} + \gamma_{p}\right]\right)^2 + \omega^2 \Gamma^2}.
\end{align}
In the propagator of the sound modes, the influence of $\gamma_H$ becomes important for $c_2^2 q^2 \sim \gamma_s\gamma_H$. This corresponds to $ \ell_H^2 \sim \lambda_T^2$. Therefore, we can safely neglect $\gamma_H$ in the propagator of the sound modes. Thus, we get
\begin{align}
    \label{eq:longitudinal_MC}
    \frac{\delta \sigma_{\mathrm{fl}}}{\sigma_0} = & \, 
    \frac{2 T}{\rho} \int  (dq) q_x^2  \int \frac{d \omega}{2\pi} 
    \frac{
\left[-\left(\gamma_k + \gamma_{\zeta}\right)
\left(\omega^2 + \gamma_s^2\right) - \gamma_s c_2^2 q^2\right]}{\left[\left(\omega^2 - c_2^2 q^2 + \gamma_s^2 -  \Gamma_s \gamma_s\right)^2 +  \omega^2 \Gamma_s^2\right]
} \nonumber \\
\times & \, \left[ \frac{
(\omega^2 + \left[\gamma_{\nu} + \gamma_{p}\right]^2) }{
\left[\left(\omega^2 - \gamma_\sigma \left[\gamma_{\nu} + \gamma_{p}\right]\right)^2 + \omega^2 \Gamma^2\right]} - \frac{(\omega^2 + \left[\gamma_{\nu} + \gamma_{p}\right]^2) + \left[\gamma_{\nu} + \gamma_{p}\right] \, \gamma_H}{\left(\omega^2 -\gamma_{\sigma} \left[\gamma_{\nu} + \gamma_{p}\right]\right)^2 + \omega^2 \left(\Gamma + \gamma_{H}\right)^2}  \right] \nonumber\\
= & \, \frac{2 T}{\rho} \int  (dq) q_x^2  \int \frac{d \omega}{2\pi} 
    \frac{\left[-\left(\gamma_k + \gamma_{\zeta}\right)
\left(\omega^2 + \gamma_s^2\right) - \gamma_s c_2^2 q^2\right]}{\left[\left(\omega^2 - c_2^2 q^2 + \gamma_s^2 -  \Gamma_s \gamma_s\right)^2 +  \omega^2 \Gamma_s^2\right]} \nonumber \\
\times & \, \left[\frac{(\omega^2 + \left[\gamma_{\nu} + \gamma_{p}\right]^2) \gamma_H \omega^2 \left( \gamma_H + 2 \Gamma \right) - \left[\gamma_{\nu} + \gamma_{p}\right] \gamma_H \left[\left(\omega^2 - \gamma_\sigma \left[\gamma_{\nu} + \gamma_{p}\right]\right)^2 + \omega^2 \Gamma^2\right]  }{
\left[\left(\omega^2 - \gamma_\sigma \left[\gamma_{\nu} + \gamma_{p}\right]\right)^2 + \omega^2 \Gamma^2\right]\left[\left(\omega^2 -\gamma_{\sigma} \left[\gamma_{\nu} + \gamma_{p}\right]\right)^2 + \omega^2 \left(\Gamma + \gamma_{H}\right)^2 \right]}\right] \nonumber \\
= & \, -\frac{2 T \gamma_H}{\rho c_2^2} \int  (dq) \frac{q_x^2}{q^2} \gamma_s \int \frac{d \omega}{2\pi} 
 \frac{\left(\Gamma + \gamma_{\sigma} + \gamma_H\right)\omega^4 + \left[\gamma_{\nu} + \gamma_{p}\right] \Gamma \left(\left[\gamma_{\nu} + \gamma_{p}\right] + \gamma_H\right)\omega^2 - \gamma_{\sigma}^2 \left[\gamma_{\nu} + \gamma_{p}\right]^3}{
\left[\left(\omega^2 - \gamma_\sigma \left[\gamma_{\nu} + \gamma_{p}\right]\right)^2 + \omega^2 \Gamma^2\right]\left[\left(\omega^2 -\gamma_{\sigma} \left[\gamma_{\nu} + \gamma_{p}\right]\right)^2 + \omega^2 \left(\Gamma + \gamma_{H}\right)^2 \right]} \nonumber \\
= & \, \frac{T \gamma_H}{\rho c_2^2} \int  (dq) \frac{q_x^2}{q^2} \frac{\gamma_s \left(\gamma_H - \gamma_{\nu} - \gamma_{p} + \gamma_{\sigma}\right)}{\left(\gamma_{\nu} + \gamma_{p} + \gamma_{\sigma}\right) \left(\gamma_H + \gamma_{\nu} + \gamma_{p} + \gamma_{\sigma}\right)
\left(\gamma_H + 2 \left[\gamma_{\nu} + \gamma_{p} + \gamma_{\sigma}\right]\right)}.
\end{align}
To get an estimate for the contribution of longitudinal fluctuations to magnetoconductivity, we set all diffusive attenuation rates to be of the same order in the above equation. Therefore,
\begin{align}
    \frac{\delta \sigma_{\mathrm{fl}}}{\sigma_0} \sim & \, \frac{T \gamma_H}{\rho c_2^2} \int  (dq) \frac{q_x^2}{q^2} \frac{1}{\gamma_{\nu}} \sim \frac{T \gamma_H}{\rho D c_2^2} \int  \frac{dq}{q}.
\end{align}
The integral diverges logarithmically. In this case, we may estimate the contribution of longitudinal fluctuations to magnetoconductivity as
\begin{align}
\label{eq:logitudinal_estimate}
    \frac{\delta \sigma_{\mathrm{fl}}}{\sigma_0} \sim & \, \frac{T \gamma_H}{\rho D c_2^2}  \ln \frac{\ell_H}{\ell_{\mathrm{ee}}},
\end{align}
Using $\gamma_H \sim D/\ell_H^2$, $c_{2} \sim v_{F}$, and the estimate for the mass density of the electron liquid in monolayer graphene at charge neutrality $\rho \sim \frac{T^3}{v^4_{F}}$, we obtain 
\begin{align}
    \frac{\delta \sigma_{\mathrm{fl}}}{\sigma_0} \sim \frac{\ell^2_{\mathrm{ee}}}{\ell_H^2} \ln \frac{\ell_H}{\ell_{\mathrm{ee}}}.
\end{align}
Setting $\ln \frac{\ell_H}{\ell_{\mathrm{ee}}} \to 1$, and comparing the coefficient of the quadratic field dependence for the transverse contribution given by Eq. \eqref{eq:result_bulk}, we obtain a parameter $\frac{\ell^2_{\mathrm{ee}}}{\ell^2_{G}}$. It can be rewritten as $\left(\gamma_p \tau_{\mathrm{ee}}\right)^2$. It shows how much the longitudinal contribution to magnetoconductivity is reduced compared to the transverse contribution. Therefore, the contribution from the longitudinal fluctuations at weak magnetic fields may be neglected.
\end{widetext}

\section{Derivation of the fluctuation contribution to the conductivity for an arbitrary aspect ratio}\label{sec:general_aspect_ratio}

\begin{widetext}
In this Appendix, we evaluate the fluctuation contribution to the effective conductivity of Hall bars of width $W$ and length $L$, accounting for the finite size effects. 
In this geometry,  Eq.~\eqref{eq:master} can be written in the following form
    \begin{align}
\label{eq:Conductivity_arbitrary_aspect_ratio}
& \, \frac{\sigma_{\rm fl}}{\sigma_0} = \frac{4 \varsigma}{\pi^2} \int_{2 \pi/L}^{{2\pi}/{\ell}} \!\! dq_x \int_{{2\pi}/{W}}^{{2\pi}/{\ell}} \!\! dq_y\; \frac{q_y^2}{\left(q_{x}^2 + q_{y}^2\right) \left(q_{x}^2 + q_{y}^2 + 4 \pi^2 r_{H}^{-2}\right)},
\end{align}
where $\ell$ is the short-distance cut-off, and $r_{H} = \frac{2 \pi}{\sqrt{\ell^{-2}_{G} + \ell^{-2}_{H}}}$ is used for brevity ($\ell_{G}$ and $\ell_{H}$ are defined in Eq. \eqref{eq:l_d_l_H_def}). Integration over $q_{y}$ yields
    \begin{align}
\label{eq:Conductivity_arbitrary_AR_integrated_over_qy}
  \frac{\sigma_{\rm fl}}{\sigma_0} = & \,
    =  \mathcal{I}_{1} - \mathcal{I}_{2} ,
\end{align}
where the definite integrals $\mathcal{I}_{2}$ and $\mathcal{I}_{2}$
are given by 
\begin{subequations}
    \label{eq:I_12_def}
\begin{align}
    \label{eq:I_1_def}
    \mathcal{I}_{1} = &\, \frac{2 \varsigma}{\pi^2} \int_{{r_{H}}/{W}}^{{r_{H}}/{\ell}}  d k^2_y \, \Bigg[\arctan \left(\frac{r_{H}}{k_{y} \ell}\right) - \arctan \left(\frac{r_{H}}{k_{y} L}\right)\Bigg] ,\\
    \label{eq:I_2_def}
    \mathcal{I}_{2} = &\, \frac{2 \varsigma}{\pi^2} \int_{{r_{H}}/{W}}^{{r_{H}}/{\ell}} d k^2_y \frac{k_{y}}{\sqrt{1 + k^2_{y}}} \Bigg[\arctan \left(\frac{r_{H}}{{\ell} \sqrt{1 + k^2_{y}}}\right) - \arctan \left(\frac{r_{H}}{L \sqrt{1 + k^2_{y}}}\right)\Bigg].
\end{align}
\end{subequations}
Here we introduced a new variable $k_{y} = \frac{q_{y} r_{H}}{2 \pi}$.

Evaluation of the first integral gives
\begin{align}
\label{eq:I1_result}
      \mathcal{I}_{1} = & \, \frac{2 \varsigma}{\pi^2} \left[\frac{r_H^2}{ \ell^2} \left(1 + \frac{\pi}{2}\right) - \frac{r^2_{H}}{W \ell} \left(1 + \frac{W}{L}\right) + \frac{r^2_{H}}{W^2} \frac{W}{L}\right] \nonumber \\
 - & \, \frac{2 \varsigma}{\pi^2} \left[\left(\frac{r_H^2}{ W^2} + \frac{r_H^2}{ \ell^2} \right) \arctan\left(\frac{W}{\ell}\right) + \left(\frac{r^{2}_{H}}{\ell^2} + \frac{r_H^2}{L^2}\right) \arctan\left(\frac{\ell}{L}\right) - \left(\frac{r^{2}_{H}}{W^2} + \frac{r^{2}_{H}}{L^2}\right) \arctan\left(\frac{W}{L}\right)\right].
\end{align}

To evaluate the integral $\mathcal{I}_{2}$ in Eq.~\eqref{eq:I_2_def} we use the identity $\arctan (z) = -\frac{i}{2} \ln{\left(\frac{1 + iz}{1 - iz}\right)}$ and perform the substitution $t = \left(\sinh{k_{y}}\right)^{-1}$
to rewrite it in the form
\begin{align}
    \mathcal{I}_{2} = & \, -i \frac{2 \varsigma}{\pi^2} \int_{t_1}^{t_2} dt \sinh^2 t \ln \left(1 + \frac{i}{\ell r_{H}^{-1} \cosh t}\right) - i \frac{2 \varsigma}{\pi^2} \int_{t_1}^{t_2} dt \sinh^2 t \ln \left(1 - \frac{i}{L r_{H}^{-1} \cosh t}\right) \nonumber \\
    + & \, i \frac{2 \varsigma}{\pi^2} \int_{t_1}^{t_2} dt \sinh^2 t \ln \left(1 - \frac{i}{\ell r_{H}^{-1} \cosh t}\right) + i \frac{2 \varsigma}{\pi^2} \int_{t_1}^{t_2} dt \sinh^2 t \ln \left(1 + \frac{i}{L r_{H}^{-1} \cosh t}\right).
\end{align}
The definite integrals in the above expression are expressed in terms of elementary functions and dilogarithms.
To write the result  a relatively compact form, we introduce the following function of two variables
\begin{align}
    F (u; c) = & \, \left(\frac{u^2}{8} - \frac{1}{8 u^2}\right) \ln \left(\frac{u + c}{u - c}\right) + \frac{c u}{4} - \frac{c^2}{8} \ln \left(\frac{u + c}{u - c}\right) - \frac{1}{4 c u} \nonumber \\
    + & \, \frac{1}{8c^2} \ln \left(\frac{u + c}{u - c}\right) + \frac{i \pi}{2} \ln u + \frac{1}{2} \Bigg[\operatorname{Li}_2 \left(-\frac{u}{c}\right) - \operatorname{Li}_2 \left(\frac{u}{c}\right)\Bigg],
\end{align}
where $\operatorname{Li}_2(z)$ is the dilogarithm function.
In this notation, we get
\begin{align}
\label{eq:I2_result}
    \mathcal{I}_{2} = & \, - i \frac{2 \varsigma}{\pi^2} \Bigg[F \left(\frac{r_{H}}{\ell}
+ \sqrt{1 + \frac{r_{H}^2}{\ell^2}}; \, i\left[\frac{r_{H}}{\ell}
- \sqrt{1 + \frac{r_{H}^2}{\ell^2}}\right]\right) - F \left(\frac{r_{H}}{W} + \sqrt{1+\frac{r_{H}^2}{W^2}}; \, i\left[\frac{r_{H}}{\ell} - \sqrt{1 + \frac{r_{H}^2}{\ell^2}}\right]\right)\Bigg] \nonumber \\
    - & \, i \frac{2 \varsigma}{\pi^2} \Bigg[F \left(\frac{r_{H}}{\ell} + \sqrt{1 + \frac{r_{H}^2}{\ell^2}}; \, -i \left[\frac{r_{H}}{L} + \sqrt{1 + \frac{r_{H}^2}{L^2}}\right]\right) - F \left(\frac{r_{H}}{W} + \sqrt{1 + \frac{r_{H}^2}{W^2}}; \, -i\left[\frac{r_{H}}{L} + \sqrt{1 + \frac{r_{H}^2}{L^2}}\right]\right)\Bigg] \nonumber \\
    + & \, i \frac{2 \varsigma}{\pi^2} \Bigg[F \left(\frac{r_{H}}{\ell} + \sqrt{1 + \frac{r_{H}^2}{\ell^2}}; \, -i\left[\frac{r_{H}}{\ell} + \sqrt{1 + \frac{r_{H}^2}{\ell^2}}\right]\right) - F \left(\frac{r_{H}}{W} + \sqrt{1+\frac{r_{H}^2}{W^2}}; \, -i\left[\frac{r_{H}}{\ell} + \sqrt{1+\frac{r_{H}^2}{\ell^2}}\right]\right)\Bigg] \nonumber \\
    + & \, i \frac{2 \varsigma}{\pi^2} \Bigg[F \left(\frac{r_{H}}{\ell} + \sqrt{1 + \frac{r_{H}^2}{\ell^2}}; \, i\left[\frac{r_{H}}{L} - \sqrt{1+\frac{r_{H}^2}{L^2}}\right]\right) - F \left(\frac{r_{H}}{W}
+ \sqrt{1+\frac{r_{H}^2}{W^2}}; \, i\left[\frac{r_{H}}{L} -\sqrt{1+\frac{r_{H}^2}{L^2}}\right]\right)\Bigg].
\end{align}

In the hydrodynamic regime, we are interested in the asymptotic form of the expressions \eqref{eq:I1_result} and \eqref{eq:I2_result} at $\ell \ll W, L$, and $\ell r_{H}^{-1} \ll 1$. In this case, the result for the conductivity simplifies to
\begin{align}
\label{eq:Main_formula_hydrodynamic_limit}
& \, \frac{\sigma_{\rm fl}}{\sigma_0}
\simeq \frac{2 \varsigma}{\pi^2} \left[\frac{r_{H}^{2}}{W L} - \frac{\pi}{2} \frac{r_{H}^{2}}{W^2} +
\left(\frac{r_{H}^{2}}{W^2} + \frac{r_{H}^{2}}{L^2}\right) \arctan \left(\frac{W}{L}\right)\right] \nonumber \\
+ & \, \frac{2 \varsigma}{\pi^2} \Bigg\{\frac{1}{2} - G + \frac{\pi}{2}
\ln \left(\frac{\frac{r_{H}}{\ell}}{\frac{r_{H}}{W} + \sqrt{1 + \frac{r^2_{H}}{W^2}}}\right) - \pi \ln \left(\frac{r_{H}}{L} + \sqrt{1 + \frac{r_{H}^2}{L^2}}
\right) \nonumber \\
+ & \, \frac{\pi}{2} \frac{r_{H}}{W} \sqrt{1 + \frac{r_{H}^2}{W^2}} - \frac{1}{4} \Bigg[\left(\frac{r_{H}}{W} + \sqrt{1 + \frac{r_{H}^2}{W^2}}\right)^2 - \left(\frac{r_{H}}{W} + \sqrt{1 + \frac{r_{H}^2}{W^2}}\right)^{-2} \nonumber \\
+ & \, \left(\frac{r_{H}}{L} + \sqrt{1 + \frac{r_{H}^2}{L^2}}\right)^2 - \left(\frac{r_{H}}{L} + \sqrt{1 + \frac{r_{H}^2}{L^2}}\right)^{-2}\Bigg] \times \arctan \left(\frac{\frac{r_{H}}{L} + \sqrt{1 + \frac{r_{H}^2}{L^2}}}{\frac{r_{H}}{W} + \sqrt{1 + \frac{r_{H}^2}{W^2}}}\right) \nonumber \\
+ & \, \frac{1}{4} \Bigg[\left(\frac{r_{H}}{W} + \sqrt{1+\frac{r_{H}^2}{W^2}}\right)^2 - \left(\frac{r_{H}}{W} + \sqrt{1+\frac{r_{H}^2}{W^2}}\right)^{-2} + \left(\frac{r_{H}}{L} + \sqrt{1 + \frac{r_{H}^2}{L^2}}
\right)^{-2} - \left(\frac{r_{H}}{L} + \sqrt{1 + \frac{r_{H}^2}{L^2}}
\right)^2\Bigg] \nonumber \\
\times & \, \arctan \left(\frac{1}{\left(\frac{r_{H}}{W} + \sqrt{1 + \frac{r_{H}^2}{W^2}}\right) \left(\frac{r_{H}}{L} + \sqrt{1 + \frac{r_{H}^2}{L^2}}\right)}\right) - \frac{1}{4} \left(\frac{r_{H}}{W} + \sqrt{1 + \frac{r_{H}^2}{W^2}}\right) \left(\frac{r_{H}}{L} + \sqrt{1 + \frac{r_{H}^2}{L^2}}\right) \nonumber \\
- & \, \frac{1}{4 \left(\frac{r_{H}}{W} + \sqrt{1 + \frac{r_{H}^2}{W^2}}\right) \left(\frac{r_{H}}{L} + \sqrt{1 + \frac{r_{H}^2}{L^2}}\right)} + \frac{1}{4} \frac{\frac{r_{H}}{W} + \sqrt{1 + \frac{r_{H}^2}{W^2}}}{\frac{r_{H}}{L} + \sqrt{1 + \frac{r_{H}^2}{L^2}}
} + \frac{1}{4} \frac{\frac{r_{H}}{L} + \sqrt{1 + \frac{r_{H}^2}{L^2}}}{\frac{r_{H}}{W} + \sqrt{1 + \frac{r_{H}^2}{W^2}}}
\nonumber \\
- & \, \operatorname{Ti}_2 \left(\frac{\frac{r_{H}}{W} + \sqrt{1 + \frac{r_{H}^2}{W^2}}}{\frac{r_{H}}{L} + \sqrt{1 + \frac{r_{H}^2}{L^2}}}\right) + \operatorname{Ti}_2 \left(\Bigg[\frac{r_{H}}{W} + \sqrt{1 + \frac{r_{H}^2}{W^2}}\Bigg] \Bigg[\frac{r_{H}}{L} + \sqrt{1+\frac{r_{H}^2}{L^2}}\Bigg]\right) \Bigg\}.
\end{align}
$G$ here is Catalan's constant, and $\operatorname{Ti}_2 (x) =\int_0^x  \frac{\arctan t }{t} \, dt  $ is the inverse tangent integral. 

Finally, introducing dimensionless variables,
\begin{align}
    x = \frac{r_{H}}{W}, \qquad y = \frac{r_{H}}{L}, \qquad z = \frac{2 r_{H}}{\ell},
\end{align}
we can express the fluctuation magnetoconductivity in the form 
\begin{align}
\label{eq:Main_formula_x_y_z}
& \, \frac{\sigma_{\rm fl}}{\sigma_0} \simeq \frac{2 \varsigma}{\pi^2} \Bigg \{xy-\frac{\pi}{2}x^2 + (x^2 + y^2)\arctan \left(\frac{y}{x}\right) + \frac{1}{2} - G + \frac{\pi}{2} \ln \left(\frac{z}{x + \sqrt{1 + x^2}}\right) - \pi \ln \left(y + \sqrt{1 + y^2}\right) \nonumber \\
+ & \, \frac{\pi x}{2} \sqrt{1 + x^2} - \frac{1}{4} \Bigg[\left(x + \sqrt{1 + x^2}\right)^2 - \left(x + \sqrt{1 + x^2}\right)^{-2} + \left(y + \sqrt{1 + y^2}\right)^2 - \left(y + \sqrt{1 + y^2}\right)^{-2}\Bigg] \times \arctan \left(\frac{y + \sqrt{1 + y^2}}{x + \sqrt{1 + x^2}}\right)
\nonumber \\
+ & \, \frac{1}{4} \Bigg[\left(x + \sqrt{1 + x^2}\right)^2 -\left(x + \sqrt{1 + x^2}\right)^{-2} + \left(y + \sqrt{1 + y^2}\right)^{-2} - \left(y + \sqrt{1 + y^2}\right)^2\Bigg] \times \arctan \left(\frac{1}{\Big[x + \sqrt{1 + x^2}\Big] \Big[y + \sqrt{1 + y^2}\Big]}\right)
\nonumber \\
- & \, \frac{1}{4} \left(x + \sqrt{1 + x^2}\right) \left(y + \sqrt{1 + y^2}\right) - \frac{1}{4 \left(x + \sqrt{1 + x^2}\right) \left(y + \sqrt{1 + y^2}\right)}
+ \frac{1}{4} \frac{x + \sqrt{1 + x^2}}{y + \sqrt{1 + y^2}} + \frac{1}{4}
\frac{y + \sqrt{1 + y^2}}{x + \sqrt{1 + x^2}} \nonumber \\
- & \operatorname{Ti}_2 \left(\frac{x + \sqrt{1 + x^2}}{y + \sqrt{1 + y^2}}\right)
+ \operatorname{Ti}_2 \left(\Big[x + \sqrt{1 + x^2}\Big] \Big[y + \sqrt{1 + y^2}\Big]\right) \Bigg\}.
\end{align}
  The expression above is the asymptotic form for $z \to \infty$, which corresponds to weak magnetic fields.
\end{widetext}


%

\end{document}